\documentclass{article}

\usepackage{PRIMEarxiv}

\usepackage[utf8]{inputenc} 
\usepackage[T1]{fontenc}    
\usepackage{hyperref}       
\usepackage{url}            
\usepackage{booktabs}       
\usepackage{amsfonts}       
\usepackage{nicefrac}       
\usepackage{microtype}      
\usepackage{lipsum}
\usepackage{fancyhdr}       
\usepackage{graphicx}       
\usepackage{amsmath}
\usepackage{comment}
\usepackage{graphicx}
\usepackage{subcaption}
\graphicspath{{media/}}     
\usepackage{authblk}
\title{Application of Ideal Observer for Thresholded Data in Search Task
}

\author{Hongwei Lin}
\author[*]{Howard C. Gifford}
\affil{Department of Biomedical Engineering, University of Houston, 3517 Cullen Blvd, Houston, TX 77204, USA}
\affil[*]{\texttt{hgifford@central.uh.edu}}

\begin{document}
\maketitle

\begin{abstract}
This study advances task-based image quality assessment by developing an anthropomorphic thresholded visual-search model observer. The model is an ideal observer for thresholded data inspired by the human visual system, allowing selective processing of high-salience features to improve discrimination performance. By filtering out irrelevant variability, the model enhances diagnostic accuracy and computational efficiency.

The observer employs a two-stage framework: candidate selection and decision-making. Using thresholded data during candidate selection refines regions of interest, while stage-specific feature processing optimizes performance. Simulations were conducted to evaluate the effects of thresholding on feature maps, candidate localization, and multi-feature scenarios. Results demonstrate that thresholding improves observer performance by excluding low-salience features, particularly in noisy environments. Intermediate thresholds often outperform no thresholding, indicating that retaining only relevant features is more effective than keeping all features.

Additionally, the model demonstrates effective training with fewer images while maintaining alignment with human performance. These findings suggest that the proposed novel framework can predict human visual search performance in clinically realistic tasks and provide solutions for model observer training with limited resources. Our novel approach has applications in other areas where human visual search and detection tasks are modeled such as in computer vision, machine learning, defense and security image analysis.
\end{abstract}


\section{Introduction}
The assessment of medical image quality plays a crucial role in evaluating and optimizing medical imaging systems\cite{barrett1993model,barrett2013foundations,kundel2006history,das2011comparison,gifford2005comparison}. Given that clinicians are the primary users of diagnostic imaging modalities, the diagnostic performance of expert human observers are considered ground truth\cite{pisano2005diagnostic,skaane2013comparison,national2011reduced}. However, large-scale studies involving radiologists during the research phase are impractical due to the significant time and effort required for medical image interpretation\cite{alexander2022mandating,kwee2021workload}. It is necessary to develop computational model observers capable of predicting expert human performance in diagnostic tasks. Significant progress has been made in the development of model observers\cite{barrett1993model,Myers:87,rolland1992effect,yao1992predicting,gifford2000channelized,lago2020foveated}, but important challenges remain. One challenge is to achieve clinical realistic in task design. This requires the task to reflect radiologists’ diagnostic activities. Also, clinical data should be considered to train, test, and validate the model.  When using clinical data, researchers will have to deal with limited training resources. To address these challenges, we propose a training-efficient model that can handle both detection and search tasks.

Widely used model observers such as the ideal observer(IO) and the channelized Hotelling observer (CHO) have provided important tools for the assessment of image quality. However, the ideal observer is computational burdensome and CHO is limited in its ability to handle location-unknown tasks\cite{shen2006using,he2008toward,myers1987addition}. 

The Visual Search Model Observer (VSMO) has been introduced to address the challenge of location-unknown tasks in medical image interpretation by simulating how radiologists search for and detect abnormalities in images using a two-stage search-and-decision process.\cite{gifford2013visual,gifford2014efficient,lau2013towards,sen2014assessment,das2011comparison,karbaschi2018assessing,das2015examining}. VSMO has been shown to be effective in various research studies. Gifford et al. conducted a comparative analysis between the VSMO and human observers on SPECT images\cite{gifford2000channelized}, demonstrating that the VSMO can predict human performance. In the study done by Lau et al.\cite{lau2013towards}, VSMO is applied to simulated DBT images for parameter optimization. Recent studies focus on how to efficiently select candidates that are most likely to draw human observer attention. Progress has been made by limiting the total number of candidates per image and setting minimum separation distances between candidates \cite {karbaschi2018assessing}. In this study, we further investigate candidate-selection strategies by implementing the ideal observer for thresholded data.

The threshold process occurs prior to the computation of feature statistics or the formation of the observer discriminant. We assess the task-based performance of existing prewhitening and nonprewhitening VSMO using original and thresholded data, comparing them against human performance. Specifically, we conduct an assessment for a location-unknown lesion-detection task formulated in a localization receiver operating characteristic (LROC) study. Our studies aim to determine whether the ideal observer applied on thresholded data is better at predicting human performance and, at the same time, reduce the required amount of training data.

\section{Background}
\subsection{Tasks}
Observers in our studies examined sets of 2D images for the presence of a lesion and estimation of the lesion location. The location-unknown tasks tested can be described by the pair of hypotheses:

\[
H_0 : f = b + n, \tag{1}
\]
\[
H_j : f = b + s_j + n, \quad (j = 1, \ldots, J) \tag{2}
\]

where \( f \) is the test image, \( b \) is the mean image background, \( n \) is the additive image noise, and \( s_j \) is the lesion centered at the \( j \)-th position (\( j = 1, \ldots, J \)) on one of \( J \) pixels. 

In the location-known task, \( J = 1 \) because one lesion is centered on one known pixel in the image. In terms of detection-location tasks, \( J \) is equal to the number of possible locations for the lesion to be centered.

\subsection{Performance Assessment}
The use of receiver operating characteristic (ROC) methods is a standard approach to evaluating observer performance in detection tasks \cite{kundel2006history}. For location-known tasks, the observer assigns to each image a numerical rating that reflects the relative belief that a lesion is present or absent. A common rating scale assigns smaller values for absence and
higher values for presence. ROC analysis plots the true-positive rate (sensitivity) against the false-positive rate (1-specificity) across thresholds. The ROC curve shows the trade-off between true and false positives, with the area under the curve (AUC) summarizing performance (0.5: guessing, 1.0: perfect).

For location-unknown tasks, observers must search the images, and localization-ROC (LROC) is used\cite{swensson1996unified}. In LROC, the true positive rate is only counted when both the signal is detected and its location is within an acceptable range of the actual signal location.
\subsection{Model observers}
\label{subsec:Model observers}
Model observers are computational tools designed to simulate the performance of human observers in diagnostic tasks. During the nascent stages of model observer development, the concept of the ideal observer is brought to the medical imaging field from signal detection theory\cite{barrett1993model,barrett2013foundations}. It was initially used to define the optimal decision-making process based on all available statistical information. In medical imaging, the ideal observer served as a theoretical benchmark for the best possible performance\cite{Myers:87}. However, the ideal observer is computationally burdensome and impractical if clinical images are used\cite{shen2006using,he2008toward}. It also failed to predict human performance in some tasks\cite{myers1985effect}. To address these challenges, researchers have widely adopted the Channelized Hotelling Observer (CHO) as a preferred model observer\cite{myers1987addition}. The CHO divides the image into different spatial frequency channels to simplify the calculation and, at the same time, approximate the human visual system\cite{burgess1989efficiency}. CHO is shown to be effective in many tasks\cite{yao1992predicting}\cite{plativsa2011channelized}\cite{gifford2000channelized}, but it is also limited to signal location-known tasks.
The visual search model observer (VSMO) provides a more advanced framework for simulating human performance in tasks where signal location is uncertain. Unlike the Channelized Hotelling observer (CHO), which is effective for tasks involving detection at known locations, the VSMO is designed to handle more complex, location-unknown search tasks\cite{gifford2013visual}.The VSMO has two stages. In the first stage, the model simulates the process by which human observers scan an image to locate potential areas of interest (e.g., lesions or anomalies), and these locations are referred to as candidates. This stage attempts to capture the global search behavior, in which human observers quickly scan an image to narrow down possible locations for closer inspection\cite{evans2013gist,wolfe2021guided}. This is an essential difference between the VSMO and employing the CHO in a scanning operation. Once potential regions of interest are identified in the search stage, the second stage involves a more focused analysis of these regions. In this stage, the VSMO applies a decision-making algorithm to determine whether a signal (such as a lesion) is present or absent\cite{gifford2014efficient}. 
\section{Methods}
\label{sec:methods}
\subsection{Data Simulation}
The lumpy-background images were used as a basic simulation of single-pinhole planar imaging \cite{Barrett1990}. The dimensions of the 2D image phantoms were 128 × 128 pixels, with a pixel size of 2.4 mm. Lesion-absent phantoms consisted of same-sized, rotationally symmetric Gaussian lumps randomly positioned on a constant background. The number of background lumps per phantom followed a Poisson distribution with a mean of 50. The full width at half maximum (FWHM) for the lumps was 28.2 pixels. Each lesion-present phantom shared the lumpy-background model of the lesion absent phantoms but also contained a 2D Gaussian function (FWHM 9.4 pixels) representing the lesion. These phantoms had the lesion inserted at randomized locations for a location-unknown lesion search task. The single-pinhole imaging process was modeled by convolving the phantom with a 2D Gaussian aperture representing the pinhole. The target-relative pinhole diameter was described as the ratio between the lesion diameter and the aperture diameter. Varying the pinhole diameter tested the trade-off between quantum noise and spatial resolution in the images. We applied relative diameters ranging from 0.2 to 3.6, with intervals of 0.2.
\subsection{Human Observer Studies}
Four observers participated in the human observer studies using a mouse-driven graphical user interface. The observers were volunteer students and faculty from the University of Houston. The studies involved both a location-known lesion detection task and a location-unknown search task.

During each session with lumpy-background images, observers underwent training with 36 images and were then tested with 72 images for relative pinhole size ratios of 0.4, 1.0, 2.0, and 3.0. Both the training and testing stages had a lesion-present image prevalence of 50\%.  Observers were required not only to select a confidence rating but also to mark the lesion position using the mouse. After inputting their decision, observers were shown the correct answer during training sessions. 

\subsection{Gabor features}
The Gabor filter was invented by Dennis Gabor, who received his Nobel Prize for the invention of the Hologram. Although Gabor’s initial work focused on one-dimensional signals, the concept was later extended to two dimensions for image processing\cite{daugman1980two}. In recent years, the Gabor filter has been widely used in edge detection, feature extraction, and texture analysis\cite{turner1986texture}\cite{caelli1985detection}. Researchers, inspired by the receptive fields of neurons in the visual cortex of mammals, employed 2D Gabor filters to simulate how biological systems detect and process visual information\cite{kamarainen2012gabor}\cite{Jones1987}.

The 2D Gabor function is defined as follows:

\begin{equation}
\begin{split}
G(x, y) = \exp\left[-4(\ln 2)\frac{(x - x_0)^2 + (y - y_0)^2}{W_s^2}\right] \\
\cdot \cos\left[2\pi f_c \left((x - x_0)\cos(\theta) + (y - y_0)\sin(\theta)\right) + \phi\right]
\end{split}\tag{3}
\end{equation}

where \( (x, y) \) are the spatial coordinates of the filter, and \( (x_0, y_0) \) represent the center of the Gabor filter in the spatial domain. The parameter \( W_s \) denotes the width of the Gaussian envelope, which controls the spatial spread of the filter. The term \( f_c \) specifies the central frequency of the sinusoidal component, determining the filter’s frequency sensitivity. The orientation of the sinusoidal wave is given by \( \theta \), allowing the filter to be tuned to specific edge directions, while \( \phi \) is the phase offset, shifting the cosine wave to match particular features in the image.

When dealing with realistic and complicated images, the ability to parameterize and customize Gabor filters makes them a versatile and powerful tool, as they can be tailored to extract specific features and patterns relevant to various tasks. In this study, the initial filter bank is built with 48 Gabor filters, varying two different \( W_s \), three different \( f_c \), four different \( \theta \), and two different \( \phi \).

\subsection{Refine feature bank}
\label{subsec:Refine feature bank}
\begin{figure}[htbp]
    \centering
    \includegraphics[width=\linewidth]{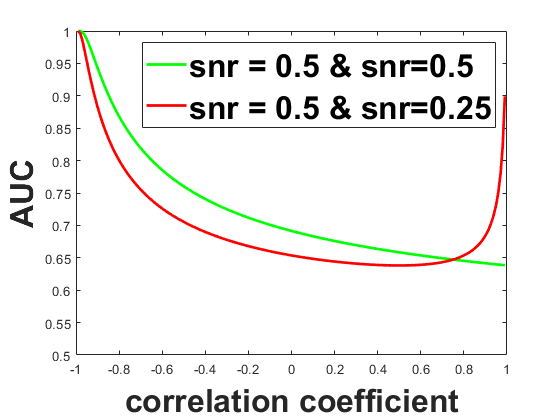} 
    \caption[ Performance comparison of 2 features.]{ Performance comparison of two features. The red plot is the combination of two features with different SNRs; the green plot is the combination of two features with same SNRs.}
    \label{fig:2 features corr}
\end{figure}
The signal-to-noise ratio (SNR) is a good metric for evaluating discriminate ability. It is calculated using equations:
\begin{equation}
\text{SNR}^2 = \Delta \mathbf{d}^\top \mathbf{K}^{-1} \Delta \mathbf{d} \label{eq:SNR},\tag{4}
\end{equation}
where:
\[
\Delta \mathbf{d} = \mathbf{d}_\text{present} - \mathbf{d}_\text{absent},
\]
and \( \mathbf{d}_\text{present} \) denote the mean feature vector for the signal-present class and \( \mathbf{d}_\text{absent} \) the mean feature vector for the signal-absent class. The covariance matrix of the feature vectors, assumed to be identical for both classes, is represented by \( \mathbf{K} \), and its inverse is denoted by \( \mathbf{K}^{-1} \).

The term \( \Delta \mathbf{d} \) represents the difference in the mean feature vectors between the two classes. The quadratic form involving \( \mathbf{K}^{-1} \) weights this difference vector by the inverse of the covariance, effectively normalizing it by the variability within the feature space. The resulting \( \text{SNR}^2 \) provides a measure of the squared discriminability between the signal-present and signal-absent classes.
Under the assumption of feature values having an underlying normal distribution, the SNR is related to AUC by the following equation:
\[
\text{SNR} = \sqrt{2} \cdot \Phi^{-1}(\text{AUC}), \tag{5}
\]
where \( \Phi \) is the error function\cite{barrett2013foundations}

Using Gaussian simulated feature values, the performances of features with the same or different SNRs are explored.
In figure~\ref{fig:2 features corr}, the result shows features with different SNRs can exceed the features with the same SNRs when correlation approaches 1. So a rule-of-thumb can be draw from these results, features that have the same SNRs have higher performance when correlation is low( < 0.3, only consider positive correlation cases for this study), and features that have different SNRs have higher performance when correlation is high(> 0.8). Using this rule, feature bank size is reduced from 48 to 12.
\subsection{Thresholding features}
The application of a threshold on the feature values introduces an augmentation to the linear discriminant process. For VSMO, the thresholding of data was implemented as an extension of the first stage to refine feature processing. Each feature was assigned an independent lower threshold, such that any feature value below this threshold was disregarded. The thresholds for different features were independent of one another. If all features for a given image fell below their respective thresholds, the image was classified as lesion-absent, and a test statistic of negative infinity was assigned. Only the feature values that surpassed their thresholds were utilized in computing the model statistics.

To illustrate, consider a case with a single feature. If the feature value from a given image exceeds the threshold, it is retained as a post-threshold value. The linear discriminant is based on the corresponding post-threshold feature values obtained from the training cases. Conversely, if a feature value does not exceed the threshold, the feature value is eliminated. In such instances, the case does not contribute to the estimation of class statistics and is assigned an extremely negative confidence rating (-9999 in code), which categorizes it as a lesion-absent case in LROC analysis.

In cases with two features, the thresholding process results in four possible outcomes: 
\begin{enumerate}
    \item Both feature values exceed the respective thresholds, and both are retained for use in the linear discriminant, which is calculated using the corresponding post-threshold features from the training cases.
    \item Only the first feature value exceeds the threshold, leading to its retention and participation in the second linear discriminant, calculated using only the first post-threshold feature from the training cases.
    \item Only the second feature values exceed the threshold, leading to a similar process where the third linear discriminant is computed using only the second post-threshold feature.
    \item Neither feature value exceeds the thresholds. In this scenario, the instance does not contribute to the estimation of class statistics and is assigned an extremely negative confidence rating, categorizing it as a lesion-absent case in LROC analysis.
\end{enumerate}
While scenarios (2) and (3) involve the same number of retained features, they represent distinct discriminant situations (3 total discriminant situations)  due to differences in the specific features retained.

 The above 2-feature example suggests the computational burden grow fast with the number of features. For a total of $n$ feature, $n^{2} - 1$ cases of class statistics need to be considered, theoretically. The computational cost is one of the reason that we manually selected a feature bank size of 12 in~\ref{subsec:Refine feature bank}.  
 
 The Thresholds were determined using training images and subsequently applied in the testing phase of the study.

\subsection{Stage-specific features}
As noted in \ref{subsec:Model observers}, the visual search model observer is a 2-stage model consisting of a search stage and a decision stage. This two-stage model reflects the cognitive and perceptual mechanisms underlying how humans detect and identify targets within a visual scene, relying on different types of features and processing modes at each stage\cite{wolfe1994guided,treisman1980feature}. At the initial search stage, the visual system relies on low-level features—such as color, orientation, spatial frequency, and motion—to detect stimuli that stand out due to their distinctiveness. Once potential targets are identified in the search stage, the process shifts to the decision stage, where the visual system engages in a focused, attentive examination of these regions\cite{treisman1980feature}. This stage is characterized by serial processing, meaning that each candidate area is analyzed individually to confirm whether it matches the target. Here, more complex, high-level features come into play, including shape, texture, and contextual cues related to the scene.\cite{treisman1980feature}

The model used in this study is designed to reflect the above principle. The candidate searching stage and decision-making stage are not fixed to use the same set of Gabor features.
This raises the question of how many features should be used Gabor features have excellent parameterization. With enough parameter variation, the Gabor filters can cover the majority of the frequency domain of the image. But this is not a "more is better" situation because according to Wolfe's research\cite{wolfe1994guided}, the human visual system has a bottleneck effect that limits the number of features that can be processed in parallel. Considering both the bottle-neck theory and the computational burden of the model, the model is set to choose 3 features from the 12 feature bank for each stage.
\subsection{Training Resource Reduction}
The effect of the number of training images on model-observer performance was explored. The training size M varied from 30, 50, 100, 200, 300, 400, and 500 images for both the lesion-present and lesion-absent cases. The total training set size was thus 2xM images. Ten trials were conducted for each training image set size. The training images were randomized in every trial while the testing images were fixed. The total testing set size was 400 images, evenly split between lesion-present and lesion-absent cases. Standard errors were calculated to quantify the uncertainty in the estimates of AUC obtained for a given training size. 
\section{Results}
\subsection{Human study results}
\begin{figure}[ht]
    \centering
    \includegraphics[width=\linewidth]{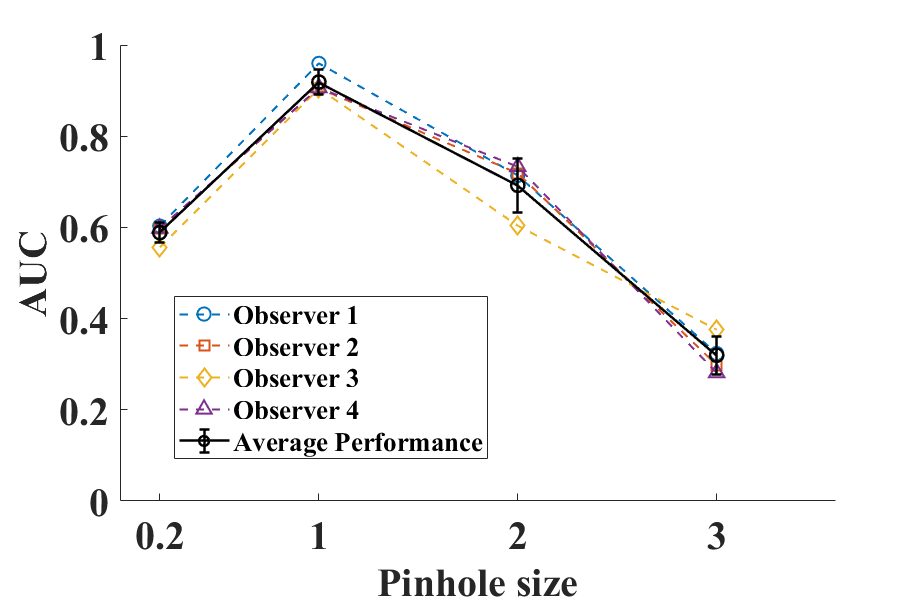} 
    \caption{ Performance of four observers in search task with lumpy-background images.}
    \label{human_lumpy_result}
\end{figure}

The human observer performance is shown in Fig.~\ref{human_lumpy_result}. The AUC starts low at smaller pinhole sizes because of limited photo count leads to higher quantum noise. The performance decreases at higher pinhole sizes due to reduced resolution and image blurring. The trade-off balanced around pinhole size ratio 1.

\subsection{Stage-dependent features}
\begin{figure}[ht]
    \centering
    \includegraphics[width=\linewidth]{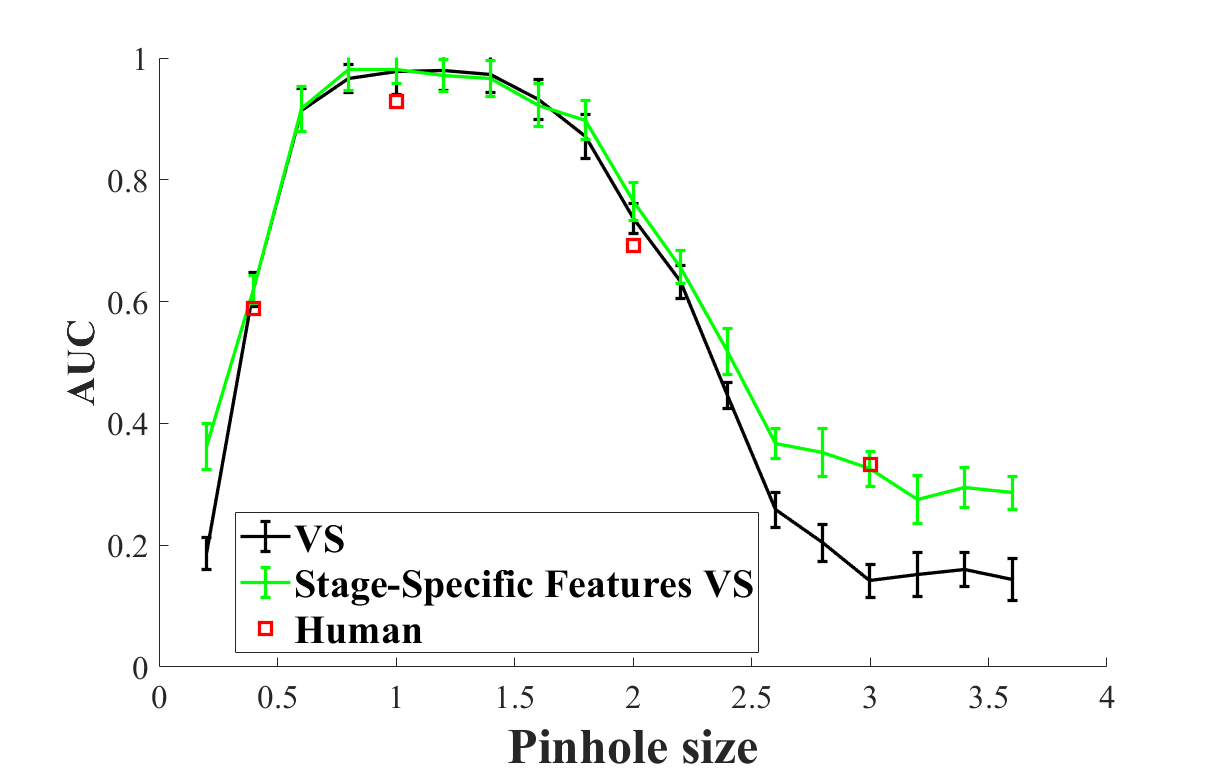} 
    \caption{ Performance comparison for the stage-specific features model (green line) and visual search observers (black line).}
    \label{2stageFeature}
\end{figure}

In Fig.~\ref{2stageFeature}, it is shown that model with different features selected at the 2 stage can perform better than the VS model. The improvement of performance is more obvious in the smaller and larger pinhole sizes. The VS model did not predict human performance accurately. The addition of using different features at different visual search stages improves the model's ability on this part.
\subsection{ Models with thresholds}
\begin{figure}[htbp]
    \centering
    
    \begin{subfigure}[b]{\columnwidth} 
        \centering
        \includegraphics[width=0.9\linewidth, keepaspectratio]{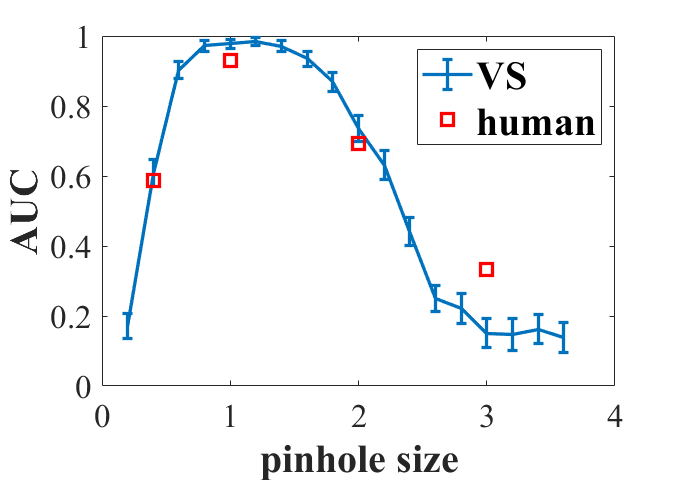} 
        \caption{Performance (AUC) of the prewhitening visual search observer (blue line) and average human observers (red squares).}
        \label{vs&human2}
    \end{subfigure}
    
    \vspace{0.5cm}
    \begin{subfigure}[b]{\columnwidth} 
        \centering
        \includegraphics[width=0.9\linewidth, keepaspectratio]{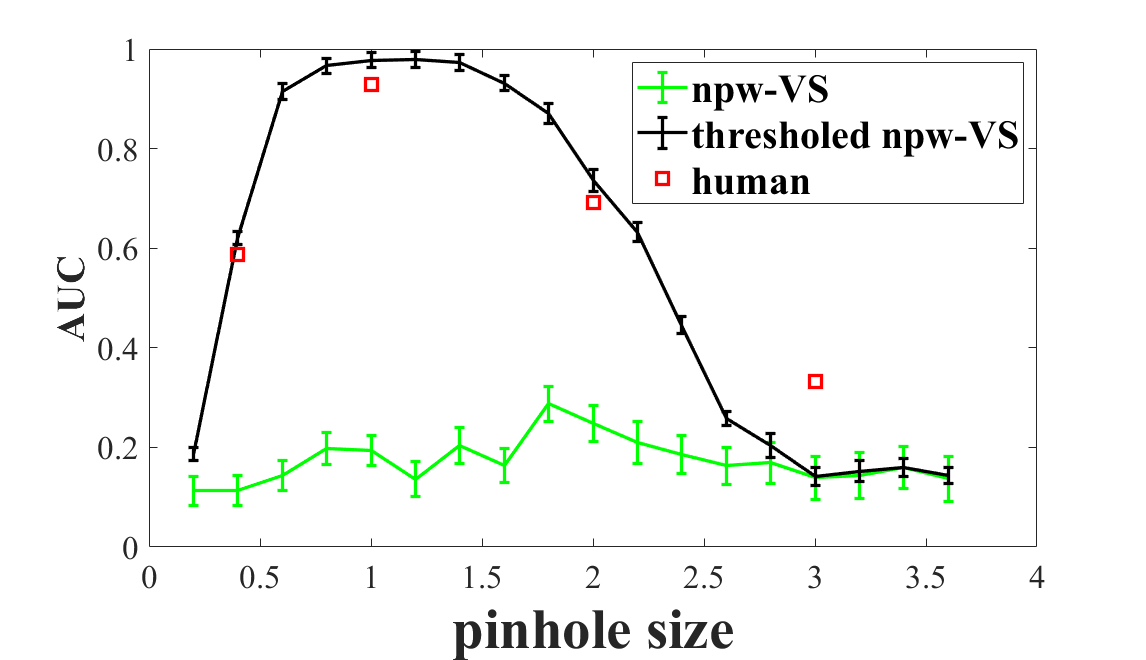} 
        \caption{Performance comparison for the nonprewhitening VS model(green line) and thresholded nonprewhitening (black line) VS model with lumpy-background images. The error bars represent +/- one standard error in the AUC measurement. The human performance is shown as
red squares.}
        \label{thrvs_training2}
    \end{subfigure}

    \caption{Perfomance comparison between prewhitening VS model, nonprewhitening VS model, thresholded nonprewhitening VS model and human.}
    \label{thr_npw}
\end{figure}

The AUC performance of the prewhitening visual search observer without thresholds is plotted as a function of pinhole diameter in Fig.~\ref{vs&human2}. The error bars represent +/- one standard error in the AUC measurement. Average human performance is also shown for comparison. The model observer predicts human performance well with the three smallest pinhole diameters. Results for the nonprewhitening visual search observer are given in Fig.~\ref{thrvs_training2}. While the nonprewhitening observer without thresholding consistently performed much worse than the prewhitening models at most pinhole diameters, the addition of thresholding to the nonprewhitening observer largely makes up the differences.

\subsection{Training resources Study}
\begin{figure}[htbp]
    \centering
    
    \begin{subfigure}[b]{\columnwidth} %
        \centering
        \includegraphics[width=0.8\linewidth, keepaspectratio]{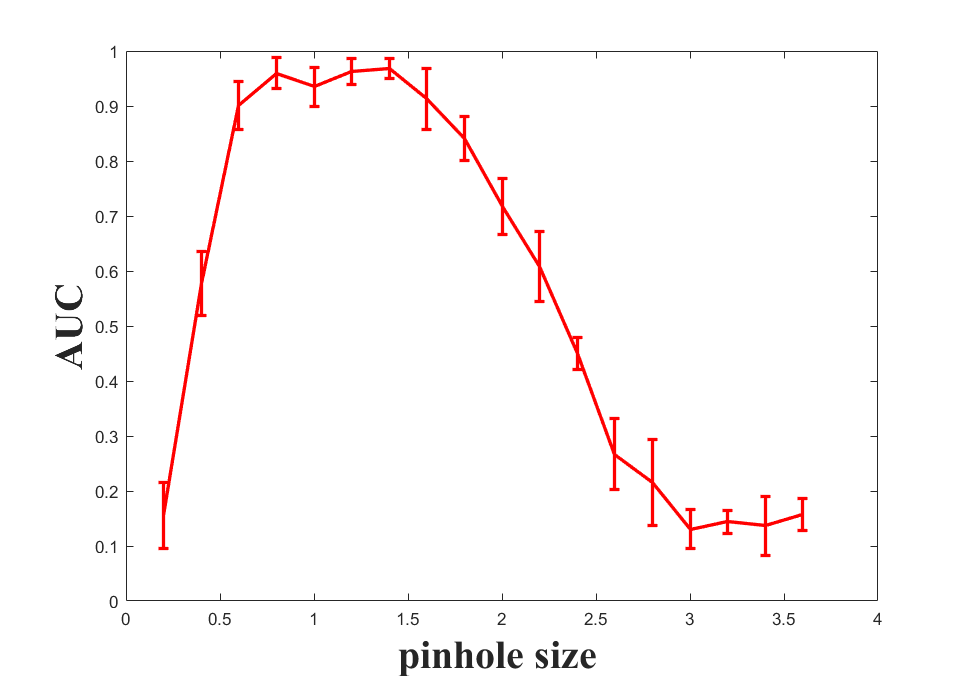} 
        \caption{Performance of the prewhitening visual search observer based on 50 training-image pairs (100 total training images).}
        \label{vs_training}
    \end{subfigure}
    
    \vspace{0.5cm}

    \begin{subfigure}[b]{\columnwidth} 
        \includegraphics[width=0.8\linewidth, keepaspectratio]{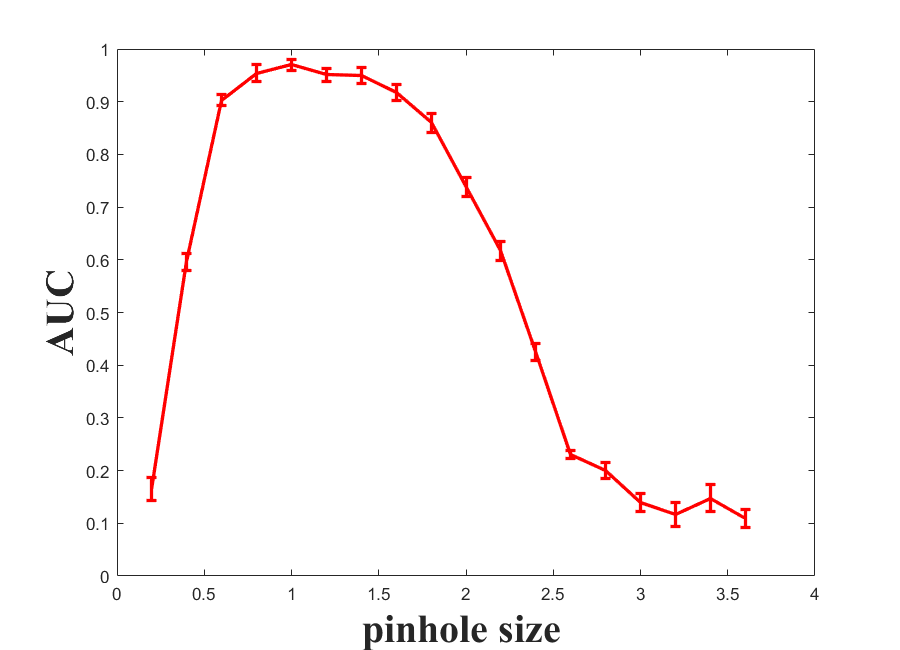} 
        \caption{Performance of the thresholded version of the model observer based on 50 training-image pairs.}
        \label{thrvs_training}
    \end{subfigure}

    \caption{Performance uncertainty comparison between the VS model, the thresholded VS model, and the human.}
    \label{thr_vs_training}
\end{figure}
\begin{figure}[ht]
    \centering
    \includegraphics[width=\linewidth]{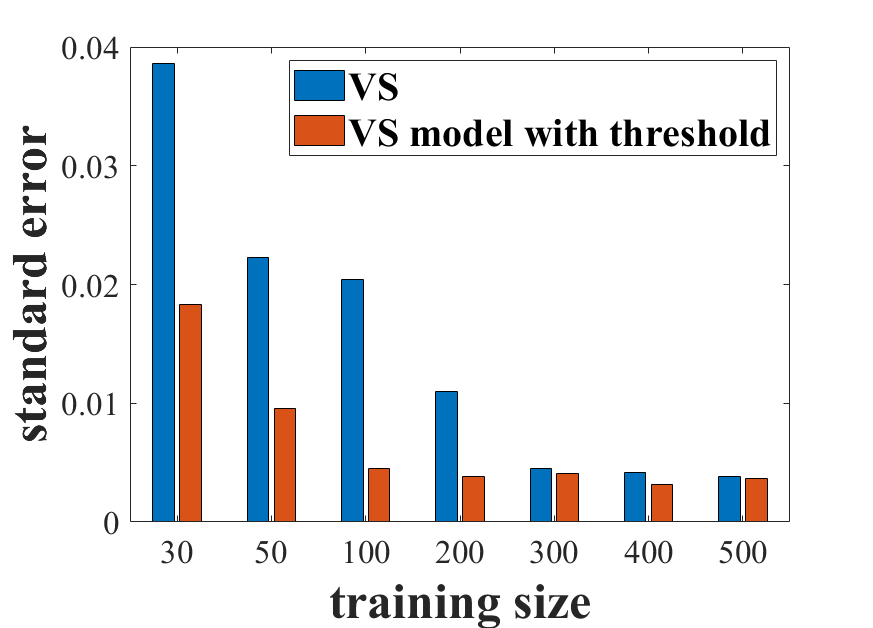} 
    \caption{ Standard errors in AUC performance for the prewhitening visual search observer with and without thresholds as a function of the number of lesion-present/lesion-absent image pairs used for training. For example, a training size of 30 represents a set of 60 training images.}
    \label{training_size_bar}
\end{figure}
The performance of the prewhitening visual search observer with and without thresholds is shown in Fig.~\ref{vs_training} and Fig.~\ref{thrvs_training}. In both cases, the observer was trained with 50 image pairs (i.e., 100 images in total). The model observer without thresholding has substantially greater performance uncertainties at all pinhole diameters compared to the thresholding model. An in-depth analysis of standard error as a function of training set size is presented in Fig.~\ref{training_size_bar}. With the smaller training sizes, thresholding led to relative reductions of greater than 50\% in standard error compared to the nonthresholded results. The standard errors for both models reach a plateau of approximately 0.005, but thresholding reaches this plateau with about one-third of the number of images needed for the nonthresholded model (200 total training cases versus 600 cases).
\section{Discussion}
\subsection{Stage-specific features}
The results show that using different feature sets for the candidate-selection and decision-making stages improves model-observer performance. Features chosen for candidate selection emphasize broad saliency cues, which restrict the later decision stage to a smaller set of plausible locations. This reduces the influence of background fluctuations and produces estimates that more closely track human search behavior at small and large pinhole sizes. The improvement relative to the baseline VSMO indicates that the selection and decision stages benefit from being tuned separately rather than sharing a single feature set.

\subsection{Effect of thresholding}
Introducing a threshold on feature values improves performance for both the prewhitening and nonprewhitening observers. Thresholding reduces the number of low-value features contributing to the discriminant, limiting the effect of noise-driven responses and reducing candidate proliferation in high-noise cases. The performance gains for the nonprewhitening model are especially notable, suggesting that thresholding compensates for the lack of prewhitening by suppressing spurious responses. Thresholding therefore provides a simple mechanism for stabilizing candidate selection in challenging settings.
\subsection{Training efficiency}
Thresholding also reduces the amount of training data needed for stable performance. For the models tested, the plateau in AUC standard error was reached with roughly one third of the training images required by the nonthresholded observers. This is consistent with the role of thresholding as a filtering mechanism that limits the influence of outliers and low-information features during parameter estimation. With fewer effective degrees of freedom, the model requires a smaller training set to estimate class statistics reliably.
\subsection{Current limitations and future work}
The spirit of using independent thresholds on each feature is that these thresholds should work as feature selection operators. The manual process of selecting and downsizing feature bank size in subsection~\ref{subsec:Refine feature bank} restricted the full potential of the thresholded model. But as of current issue of computational burden, the feature bank size has to be set at 12. In the near future, effort on GPU computing are being made to lift the restriction on total feature number. 

Research is also underway to expand the feature types from Gabor functions to texture features. An eye-tracking study is being prepared to seek correlation between human fixations and texture features. 

We are also preparing simulated and clinical digital breast tomosythesis data to generalize the current model.
\section{Conclusion}
This work evaluated an idea observer for thresholded data in a location-unknown lesion-search task. Thresholded data was used in the candidate-selection stage and tested with both prewhitening and nonprewhitening observers. The thresholded data improved observer performance, most clearly for the nonprewhitening model, and reduced the amount of training data needed for stable estimation. The study also showed that assigning different feature sets to the candidates selection stage and decision-making stage yields improved agreement with human-observer trends across imaging conditions. These results indicate that thresholding and stage-specific feature design are practical additions for developing training-efficient model observers for clinically realistic search tasks.

\section*{Acknowledgments}
This work is supported in part by R01EB032416.

\bibliographystyle{unsrt}  
\bibliography{references}

\end{document}